\documentclass[showpacs,preprintnumbers,amsmath,amssymb,floatfix,elsart,prd]{revtex4}

\usepackage{graphicx}
\usepackage{dcolumn}
\usepackage{bm}

\newcommand{\beq}{\begin{equation}}
\newcommand{\eeq}{\end{equation}}
\newcommand{\bey}{\begin{eqnarray}}
\newcommand{\eey}{\end{eqnarray}}

\begin{document}

\title{The Finslerian compact star model}

\author{Farook Rahaman}
\email{rahaman@iucaa.ernet.in} \affiliation{Department of
Mathematics, Jadavpur University, Kolkata 700032, West Bengal,
India}

\author{Nupur Paul}
\email{nnupurpaul@gmail.com} \affiliation{Department of
Mathematics, Jadavpur University, Kolkata 700032, West Bengal,
India}

\author{S.S. De}
\email{ ssddadai08@rediffmail.com} \affiliation{Department of
Applied Mathematics, University of Calcutta, Kolkata 700009,
India}

\author{Saibal Ray}
\email{saibal@iucaa.ernet.in} \affiliation{Department of Physics,
Government College of Engineering \& Ceramic Technology, Kolkata
700010, West Bengal, India}

\author{Md. Abdul Kayum Jafry}
\email{akjafry@yahoo.com} \affiliation{Department of Physics,
Shibpur Dinobundhoo Institution, Howrah 711102, West Bengal,
India}

\date{\today}

\begin{abstract}
We construct a toy model for compact stars based on the Finslerian
structure of spacetime. By assuming a particular mass function, we
find an exact solution of the Finsler-Einstein field equations
with an anisotropic matter distribution. The solutions are
revealed to be physically interesting and pertinent for the
explanation of compact stars.
\end{abstract}


\maketitle

\section{INTRODUCTION}

Spherically symmetric spacetime related astrophysical problems
have been always interesting to mathematician as well as
physicists. This is because of the fact that phenomena such as
black holes, wormholes and compact stars (starting from dwarf
stars via neutron stars to quark/strange stars through the
vigorous process of gravitational collapse) have been originated
in the class of system with spherical symmetry.

After the monumental construction of Einstein's general theory of
relativity in the period 1907 to 1915 \cite{Pais1982}, numerous
investigators have been started studying relativistic stellar
models with various aspects of physical reality. The investigation
of the exact solutions recounting static isotropic and anisotropic
astrophysical objects has continuously fascinated scientists with
growing interest and attraction. However, it has till now been
observed that most of the exact interior solutions (both isotropic
and anisotropic) of the gravitational field equations do not
fulfill in general the required physical conditions of the stellar
systems.

The existence of massive compact stellar system was first proposed
by Baade and Zwicky in 1934 \cite{BZ1934} when they argued that
supernova may yield a very small and dense star consisting
primarily of neutrons. It eventually came in to reality by the
discovery of pulsar, a highly magnetized and rotating neutron
star, in 1967 by Bell and Hewish \cite{Longair1994,Ghosh2007}.
There after, the theoretical investigation of compact stars became
fundamental area of importance in astrophysics. However, for
modeling a compact star emphasis has been given in general on the
homogeneity of the spherically symmetric matter distribution and
thus assumption was always valid for the perfect fluid obeying
Tolman-Oppenheimer-Volkoff (TOV) equation.

It was Ruderman \cite{Ruderman1972} who first argued that the
nuclear matter density ($\rho \sim 10^{15}$~gm/cc), which is
expected at the core of the compact terrestrial object, becomes
very much anisotropic. In such case of anisotropy the pressure
inside the fluid sphere can specifically be decomposed into two
parts: radial pressure and the transverse pressure, where they are
orthogonal to each other. Therefore it is quite reasonable to
consider pressure anisotropy in any compact stellar model. In this
context it can be noted that Gokhroo and Mehra \cite{gokhroo1994}
have shown that in case of anisotropic fluid the existence of
repulsive force helps to construct compact objects.

Other than the above mentioned ultra density \cite{Ruderman1972}
anisotropy may occur for different reasons in the compact stellar
system. Kippenhahn and Weigert \cite{Kippenhahn1990} have argued
that anisotropy could be introduced due to the existence of solid
core or for the presence of type $3A$-superfluid. Some other
reasonable causes for arising anisotropy are: different kind of
phase transitions \cite{sokolov1980}, pion condensation
~\cite{sawyer1972}, effects of slow rotation in a star
\cite{Silva2014} etc. However, Bowers and Liang \cite{Bowers1974}
indicated that anisotropy might have non-negligible effects on
such parameters like equilibrium mass and surface redshift. In
connection to pressure anisotropy inside a compact star several
recent theoretical investigations are available in the literature
\cite{Herrera2004,Varela2010,Rahaman2010a,Rahaman,Rahaman2012a,Kalam2012,Hossein2012,Kalam2013}.
However, there is an exhaustive review on the subject of
anisotropic fluids by Herrera and Santos \cite{Herrera1997} which
provides almost all references until 1997 and hence may be looked
at for further information.

Several major characteristics of compact stars established by the
present day observations have been tackled by Einstein's general
theory of relativity based on Riemannian geometry. Ever since the
beginning of the general theory of relativity, there has also been
considerable interest in Alternative theories of gravitation. One
of the most stimulating alterations of general relativity is that
proposed by Finsler \cite{Bao2000}.

 The first self-consistent Finsler geometry model was studied by E. Cartan  \cite{Car}   in 1935 and the
Einstein-Finsler equations for the Cartan d-connection where introduced in 1950, see  \cite{hor}. Latter,
there were studied various models of Finsler geometry and certain applications physics, see \cite{Bao2000,vac1}.
The first problem of those original works is that certain Finsler connections (due Chern-Rund
and/or Berwald) were with nonmetricity fields, see details and critics in \cite{vac1,vac2}.  The second and
third conceptual and technical problems are related to the facts that the geometric construc-
tions were in the bulk for local Euclidean signatures. In some cases, Finsler pseudo-Riemannian
configurations were considered but researchers were not able to find any exact solution.
   In a self-consistent manner and related to standard theories, relativistic models of Finsler gravity
and generalizations were constructed beginning 1996, see \cite{vac3,vac4}, when Finsler gravity and locally
anisotropic spinors were derived in low energy limits of superstring/ supergravity theories with
N-connection structure (velocity type coordinates being treated as extra-dimensional ones).
Using Finsler geometry methods, it was elaborated the so-called anholonomic frame deformation
method, AFDM, which allows to construct generic off-diagonal exact solutions in various modified gravity theories, including various commutative and noncommutative Finsler generalizations,
and in GR, see \cite{vac5,vac6,vac7}. This way various classes of exact solutions for Finsler modifications
of black hole, black ellipsoid / torus/ brane and string configurations, locally anisotropic cosmological solutions have been constructed for the so-called canonical d-connection and Cartan d-connections.

 The Finslerian space is very
suitable for studying anisotropic nature of spacetime (it's
mathematical aspects can be obtained in detail in Sec. II).
Basically this space is a generalization of Riemannian space and
has been studied in several past years extensively in connection
to astrophysical problems, e.g. L{\"a}mmerzahl et al.
\cite{Lammerzahl2012} have investigated observable effects in a
class of spherically symmetric static Finslerian spacetime whereas
Pavlov \cite{Pavlov2010} searches for applicable character of the
Finslerian spacetime by raises the question ``Could kinematical
effects in the CMB prove Finsler character of the space-time?''
Another astrophysics oriented application of the Finslerian
spacetime can be noted through the work of Vacaru
\cite{Vacaru2010} where the author has studied Finsler black holes
induced by noncommutative anholonomic distributions in Einstein
gravity.

Therefore, in the present investigation our sole aim is to
construct a toy model for compact star under the Finslerian
spacetime which can provide justification of several physical
features of the stellar system. The outline of the study is as
follows: In Sec. II the basic equations based on the formalism of
the Finslerian geometry are discussed whereas a set of specific solutions
for compact star under Finslerian spacetime has been produced in
Sec. III. The exterior spacetime and junction conditions are
sought for in Sec. IV in connection to certain observed compact
stars. In Sec. V, through several Subsections, we discuss in a
length various physical properties of the model. We pass some
concluding remarks in Sec. V for the status of the present model
as well as future plans of the work to be pursued.

\section{The basic equations based on the formalism of the Finslerian geometry}

Usually, Finslerian geometry can be constructed from the so called
Finsler structure $F$ which obeys the property
\[   (x, \mu y) = \mu F (x, y) \] for all $ \mu > 0$, where $ x
\in   M $ represents position and $ y = \frac{dx }{dt}$ represents
velocity. The Finslerian metric is given as \cite{Li2014}
\begin{equation}
g_{\mu\nu}\equiv\frac{\partial}{\partial
y^\mu}\frac{\partial}{\partial y^\nu}\left(\frac{1}{2}F^2\right)
\end{equation}

It is to be noted here that a Finslerian metric coincides with
Riemannian, if $F^2$ is a quadratic function of $y$.

The standard  geodesic equation in Finsler manifold can be
expressed as,
\begin{equation}
\frac{d^2x^\mu}{d\tau^2}+2G^\mu=0,
\end{equation}
where
\begin{equation}
G^\mu=\frac{1}{4}g^{\mu\nu}\left(\frac{\partial^2F^2}{\partial
x^\lambda\partial y^\nu}y^\lambda-\frac{\partial F^2}{\partial
x^\nu}\right),
\end{equation}
is called geodesic spray coefficients. The geodesic equation (2)
indicates that the Finslerian structure F is constant along the
geodesic.

The invariant quantity, Ricci scalar in Finsler geometry is given
as
\begin{equation}
Ric\equiv R^\mu_\mu=\frac{1}{F^2}\left(2\frac{\partial
G^\mu}{\partial x^\mu}-y^\lambda \frac{\partial^2G^\mu}{\partial
x^\lambda\partial y^\mu}+2G^\lambda\frac{\partial^2G^\mu}{\partial
y^\lambda\partial y^\mu}-\frac{\partial G^\mu}{\partial
y^\lambda}\frac{\partial G^\lambda}{\partial y^\mu}\right),
\end{equation}
where $R^\mu_\nu=R^\mu_{\lambda\nu\rho}y^\lambda y^\rho/F^2$.\\

Here, $R^\mu_{\lambda\nu\rho}$ depends on connections whereas
$R^\mu_\mu$ does not rather it depends only on the Finsler
structure F and is insensitive to connections.

Let us consider the Finsler structure is of the form \cite{Li2014}
\begin{equation}
F^2=B(r)y^ty^t-A(r)y^ry^r-r^2\bar{F}^2(\theta,\varphi,y^\theta,y^\varphi).
\end{equation}
Then, the Finsler metric can be obtained  as
\begin{equation}
g_{\mu\nu}=diag(B,-A,-r^2\bar{g_{ij}})
\end{equation}
\begin{equation}
g^{\mu\nu}=diag(B^{-1},-A^{-1},-r^2\bar{g}^{ij})
\end{equation}
where the metric   $\bar{g}_{ij}$ and its reverse are derived from
$\bar{F}$ and
 the index i,j run over angular coordinate $\theta,\phi$.

Substituting the Finsler structure (5) into Eq. (3), we find
\begin{equation}
G^t=\frac{B'}{2B}y^ty^r
\end{equation}
\begin{equation}
G^r=\frac{A'}{4A}y^ry^r+\frac{B'}{4A}y^ty^t-\frac{r}{2A}\bar{F^2}
\end{equation}
\begin{equation}
G^\theta=\frac{1}{r}y^\theta y^r+\bar{G^\theta}
\end{equation}
\begin{equation}
G^\phi=\frac{1}{r}y^\phi y^r+\bar{G^\phi}
\end{equation}
where the prime denotes the derivative with respect to $r$, and
the $\bar{G}$ is the gaodesic spray coefficients derived by
$\bar{F}$. Plugging the geodesic coefficient (8),~(9),~(10) and
(11) into the formula of Ricci scaler (4), we obtain
\begin{equation}
F^2Ric=\left[\frac{B''}{2A}-\frac{B'}{4A}\left(\frac{A'}{A}+\frac{B'}{B}\right)+\frac{B'}{rA}\right]y^ty^t
+\left[-\frac{B''}{2B}+\frac{B'}{4B}\left(\frac{A'}{A}+\frac{B'}{B}\right)+\frac{A'}{rA}\right]y^ry^r
+\left[\bar{R}ic-\frac{1}{A}+\frac{r}{2A}\left(\frac{A'}{A}-\frac{B'}{B}\right)\right]\bar{F}^2,
\end{equation}
where $\bar{R}ic$ denotes the Ricci scalar of Finsler structure
$\bar{F}$.

Now, we are in a position  to write the  self-consistent
gravitational field equation in Finsler spacetime. In a pioneering
work Akbar-Zadeh \cite{Akbar1988} first introduced the notion of
Ricci tensor in Finsler geometry as
\begin{equation}
Ric_{\mu\nu}=\frac{\partial^2(\frac{1}{2}F^2Ric)}{\partial
y^\mu\partial y^\nu}.
\end{equation}

Here the scalar curvature in Finsler geometry is defined as
$S=g^{\mu\nu}Ric_{\mu\nu}$. Therefore, the modified Einstein
tensor in Finsler spacetime takes the following form as
\begin{equation}
G_{\mu\nu}\equiv Ric_{\mu\nu}-\frac{1}{2}g_{\mu\nu}S
\end{equation}

Using equation of Ricci scalar (12), one can obtain from (13), the
Ricci tensors in Finsler geometry. This immediately yield the
Einstein tensors in Finsler geometry (note that $\bar{F}$ is two
dimensional Finsler spacetime with constant flag curvature
$\lambda$ ) as follows:
\begin{equation}
G^t_t=\frac{A'}{rA^2}-\frac{1}{r^2A}+\frac{\lambda}{r^2},
\end{equation}

\begin{equation}
G^r_r=-\frac{B'}{rAB}-\frac{1}{r^2A}+\frac{\lambda}{r^2},
\end{equation}

\begin{equation}
G^\theta_\theta=G^\phi_\phi=-\frac{B''}{2AB}-\frac{B'}{2rAB}+\frac{A'}{2rA^2}+\frac{B'}{4AB}\left(\frac{A'}{A}+\frac{B'}{B}\right).
\end{equation}

It has been shown by Li and Chang \cite{Li2014} that the covariant
derivative of Einstein tensors in Finsler geometry $G_{\mu \nu }$
vanishes i.e. covariant conserve properties of the tensor $G_{\mu
\nu }$ indeed satisfy.

Following the notion of general relativity, one can write
gravitational field equations in the given Finsler spacetime as ( \textbf{see the appendix for the justification} )
\begin{equation}
G^\mu_\nu=8\pi_FGT^\mu_\nu,
\end{equation}
where $T^\mu_\nu$ is the energy-momentum tensor.

Note that the volume of Riemannian geometry is not equal to that
of Finsler space, therefore, it is safe to use $4\pi_F$ for
expressing the volume of $\bar{F}$ in the field equation.

The matter distribution of a compact star is still a challenging
issue to the physicists, therefore, as our target to find the
interior of a compact star we assume the general anisotropic
energy-momentum tensor \cite{Rahaman2010} as
\begin{equation}
 T_\nu^\mu=(\rho + p_r)u^{\mu}u_{\nu} + p_r g^{\mu}_{\nu}+
            (p_r -p_t )\eta^{\mu}\eta_{\nu},
\end{equation}
where $u^{\mu}u_{\mu} = - \eta^{\mu}\eta_{\mu} = 1$, $p_t$ and
$p_r$ are transverse and radial pressures, respectively.

Using the above energy momentum tensor (19), one can write the
gravitational field equations in Riemannian geometry as
\begin{equation}
8\pi_F G \rho=\frac{A'}{rA^2}-\frac{1}{r^2A}+\frac{\lambda}{r^2},
\end{equation}

\begin{equation}
-8\pi_F Gp_r=-\frac{B'}{rAB}-\frac{1}{r^2A}+\frac{\lambda}{r^2},
\end{equation}

\begin{equation}
-8\pi_F
Gp_t=-\frac{B''}{2AB}-\frac{B'}{2rAB}+\frac{A'}{2rA^2}+\frac{B'}{4AB}\left(\frac{A'}{A}+\frac{B'}{B}\right).
\end{equation}

Using Eq. (23) we get the value of $A$, which is given below as
\begin{equation}
A^{-1}=\lambda-\frac{2Gm(r)}{r},
\end{equation}
where $m(r)$ is the mass contained in a sphere of radius $r$
defined by
\begin{equation}
m'(r)=4\pi_Fr^2\rho.
\end{equation}

\section{The model solution for compact star under the Finslerian spacetime}

To construct a physically viable model as well as to  make the
above set of equations solvable, we choose the mass function
$m(r)$ in a particular form that has been considered by several
authors for studying isotropic fluid spheres \cite{Finch1989},
dark energy stars \cite{Lobo2006}, anisotropic stars
\cite{Mak2003,Sharma2007} as
\begin{equation}
m(r)=\frac{br^3}{2\left(1+ar^2\right)},
\end{equation}
where two constants $a,~b$ are positive. The motivation of this
particular choice of mass function lies on the fact that it
represents a monotonic decreasing energy density in the interior
of the star. Also it gives the energy density to be finite at the
origin $r = 0$. The constants may be determined from the boundary
conditions.

Putting the value of $m(r)$ in Eq. (24), we get
\begin{equation}
\rho=\frac{b(3+ar^2)}{8\pi_F(1+ar^2)}.
\end{equation}

To determine the unknown metric potentials and physical
parameters, we use the usual equation of state
\begin{equation}
p_r=\omega \rho,
\end{equation}
where the equation of state parameter $\omega$ has the constrain
$0<\omega<1$.

Usually, this equation is used for a spatially homogeneous cosmic
fluid, however, it can be extended to inhomogeneous spherically
symmetric spacetime, by assuming that the radial pressure follows
the equation of state $p_r=\omega \rho$ and the transverse
pressure may be obtained from field equations.

Plugging Eqs. (26) and (27) in the fields Eqs. (20)-(23), we get
the explicit expressions of the unknowns in the following forms:
\begin{equation}
A=\frac{1+ar^2}{\lambda+a\lambda r^2-Gbr^2},
\end{equation}

\begin{equation}
\frac{B}{B_0}=(1+ar^2)^\omega(-\lambda-a\lambda
r^2+Gbr^2)^{\left[\frac{b-2\omega
a\lambda+3Gb\omega}{2(a\lambda-Gb)}\right]},
\end{equation}
where $B_0$ is an integration constant and without any loss of
generality, one can take it as unity.

The radial and tangential pressures are given by
\begin{equation}
p_r=\omega\rho=\frac{\omega}{8\pi_F}\left[\frac{b(3+ar^2)}{(1+ar^2)^2}\right],
\end{equation}

\begin{equation}p_t=\frac{4\omega b\lambda(3+2ar^2) -2 \omega abr^4(2 \omega \lambda-7G\omega b-2Gb-3Gb\omega^2)+Ga^2b^2r^6(1+\omega^2+2\omega)+3Gb^2r^2(1+3\omega^2)
}{32\pi_F (\lambda+a\lambda r^2-Gbr^2) (1+ar^2)^3}.
\end{equation}

Note from the above expressions for radial and tangential
pressures that the solutions obtained here are regular at the
center. Now, the central density can be obtained as
\begin{equation}
\rho(r=0)= \frac {3 b}{8\pi_F}.
\end{equation}

The anisotropy of pressures dies out at the center and hence we
have
\begin{equation}
p_r(r=0) = p_t(r=0)= \frac {3 \omega b }{8\pi_F}.
\end{equation}
 One can notice that as we match our interior solution with
external vacuum solution (pressure zero ) at the boundary, then, at
the boundary, all the components of the physical parameters are
continuous along the tangential direction ( i.e. zero pressure ),
but in normal direction it may not be continuous. Therefore, at the
boundary, pressure  may zero along tangential direction, but in
normal direction it may not be zero. So, non zero pressure at the
boundary is not unrealistic.

\section{Exterior Spacetime and Junction Condition}

Now, we match the interior spacetime to the exterior vacuum
solution at the surface with the junction radius $R$. The exterior
vacuum spacetime in Finslerian spacetime is given by the metric
\cite{Li2014}
\begin{equation}
F^{2}=\left(1-\frac{2MG}{\lambda
r}\right)y^ty^t-\left(\frac{1}{\lambda-\frac{2MG}{r}}\right)y^ry^r+r^{2}\bar{F}^2(\theta,\varphi,y^\theta,y^\varphi).
\end{equation}

Across the boundary surface $r= R$ between the interior and the
exterior regions of the star, the metric coefficients $g_{tt}$ and
$g_{rr}$ both are continuous. This yields the following results:
\begin{equation}
\lambda- \frac{2MG}{ R}   = \frac{1+aR^2}{\lambda+a\lambda
R^2-GbR^2},
\end{equation}

\begin{equation}
1 - \frac{2MG}{\lambda R} =  (1+aR^2)^\omega(-\lambda-a\lambda
R^2+GbR^2)^{\left[\frac{b-2\omega
a\lambda+3Gb\omega}{2(a\lambda-Gb)}\right]}.
\end{equation}

The above two equations contain four unknown quantities, viz.,
$a,~b,~\lambda, ~\omega$. Equation (32) yields the unknown $b$ in
terms of central density. Also from the total mass of star $m(r=R)
=M =\frac{bR^3}{2\left(1+aR^2\right)} $, we can find out the
constant $a$ in terms of the total mass $M$, radius $R$ and central
density. Finally, Eqs. (35) and (36) yield the unknowns - the flag
curvature $\lambda$ and the equation of state parameter $\omega$ in
terms of the total mass $M$, radius $R$ and central density. The
values of the constants $a,~b$ for different strange star candidates
are given in Table 1.  Note that for matching we have used
four constraint equations with four unknown and all the unknown
parameters are found in terms of R and M. For the use of continuity
of $ \frac{d g_{tt} }{d r}$, we will get an extra equation which
gives a restriction   equation  of M and R . As we have used real
parameters of mass and radius of different compact stars like PSR
J1614-2230  etc, we avoid this continuity   of  $\frac{d g_{tt }}{d
r}$.

\begin{table}
\caption{Values of the constants $a,~b$ for different strange star
candidates.} {\begin{tabular}{@{}cccccc@{}} \toprule Strange star
candidate & $R$ & $M$  & $M$ & $b$& $a$
\\ & (in km) & (in $M_\odot$) & (in km)& & \\
\colrule PSR J1614-2230 &  10.3&  1.97 & 2.905 &
0.0175& 0.0216\\ Vela X-12 &  9.99 & 1.77  & 2.610 & 0.0170&
0.0225\\ PSR J1903+327 &  9.82 & 1.67  & 2.458 & 0.0165& 0.0226\\
Cen X-1 & 9.51 & 1.49  & 2.197 & 0.0160& 0.0236\\ SMC X-4 &  8.9 &
1.29  & 1.902 & 0.015& 0.0237\\\\ \botrule
\end{tabular}}
\end{table}

\section{Physical features of the compact star model}

\subsection{Mass-radius relation}

The study of redshift of light emitted at the surface of the
compact objects is important to get observational evidence of
anisotropies in the internal pressure distribution. Before,
finding out the redshift, we give our attention to the
basic requirement of the model that whether matter distribution
will follow the Buchdahl \cite{Buchdahl1959} maximum allowable
mass-radius ratio limit. We have already found the  mass of the
star which has been given in Eq. (21).

The compactness of the star can be expressed as
\begin{equation}
u=\frac{m(r)}{r}= \frac{br^2}{2(1+ar^2)},
\end{equation}
and the  the corresponding surface redshift can be obtained as
\begin{equation}
Z_s=(1-2u)^{-\frac{1}{2}}-1=\left(1-\frac{br^2}{1+ar^2}\right)^{-\frac{1}{2}}-1.
\end{equation}

The variation of mass, compactness factor and redshift are shown
in Fig.~1 for different strange star candidates for a fixed value
of $\lambda =0.01$ whereas the maximum  mass, compactness factor
and redshift are shown in Table 2.

We have found out that
\[\left(u=\frac{m(r)}{r}\right)_{max} < \frac{4}{9}.\] Therefore,
one can note that Buchdahl's limit (which is equivalent to
$Z_s \leq 2$, the upper bound of a compressible fluid star) has
been satisfied in our model and hence it is physically acceptable.

The surface redshift $Z_s$ can be measured from the X-ray spectrum
which gives the compactness of the star. In our study, the high
redshift ($0.36-0.49$) are consistent with strange stars which
have mass-radius ratio higher than neutron stars ($Z_s
\leq 0.9$) \cite{Lindblom1984}.

\begin{figure*}[thbp]
\begin{tabular}{rl}
\includegraphics[width=5.0cm]{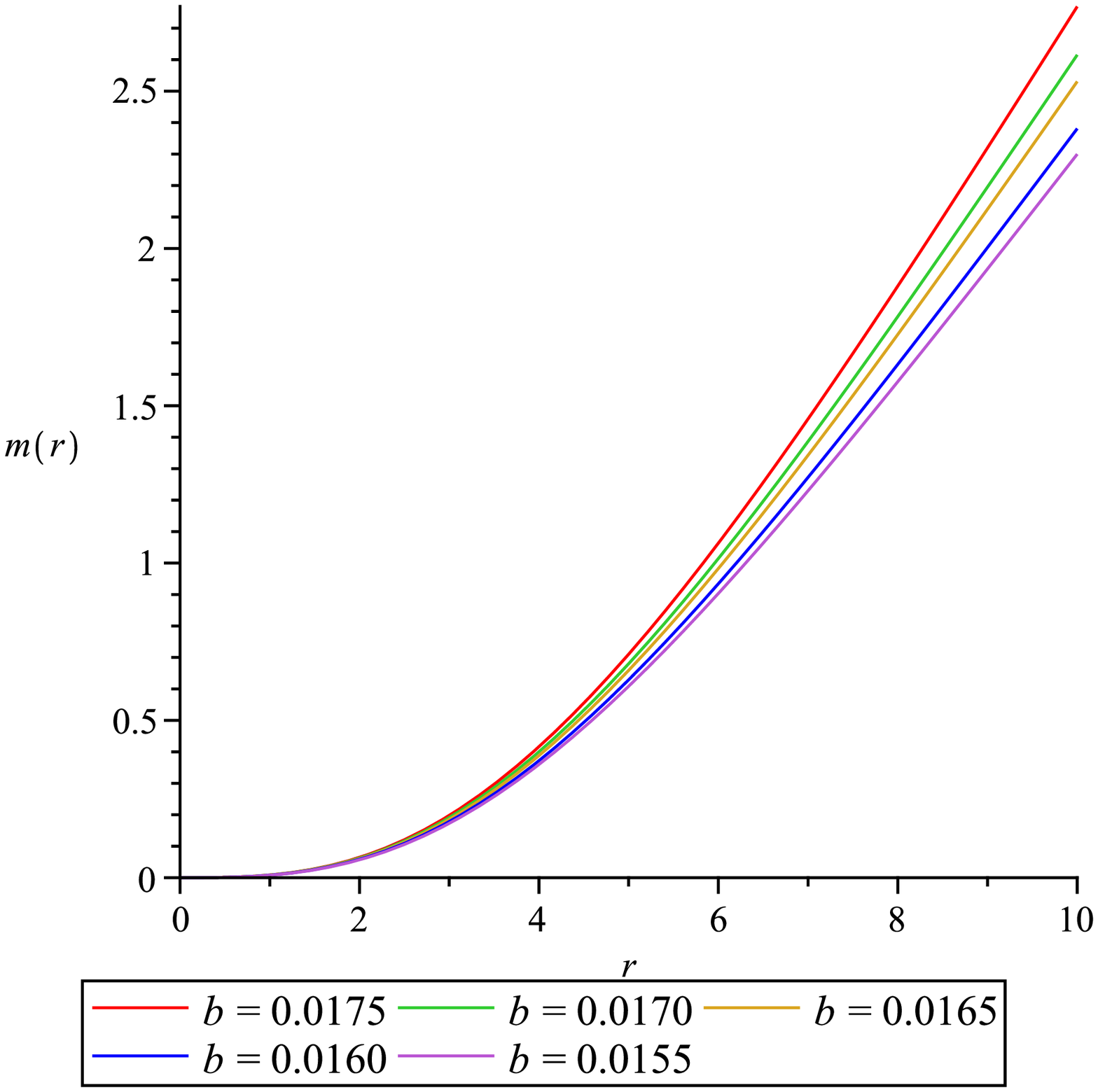}&
\includegraphics[width=5.0cm]{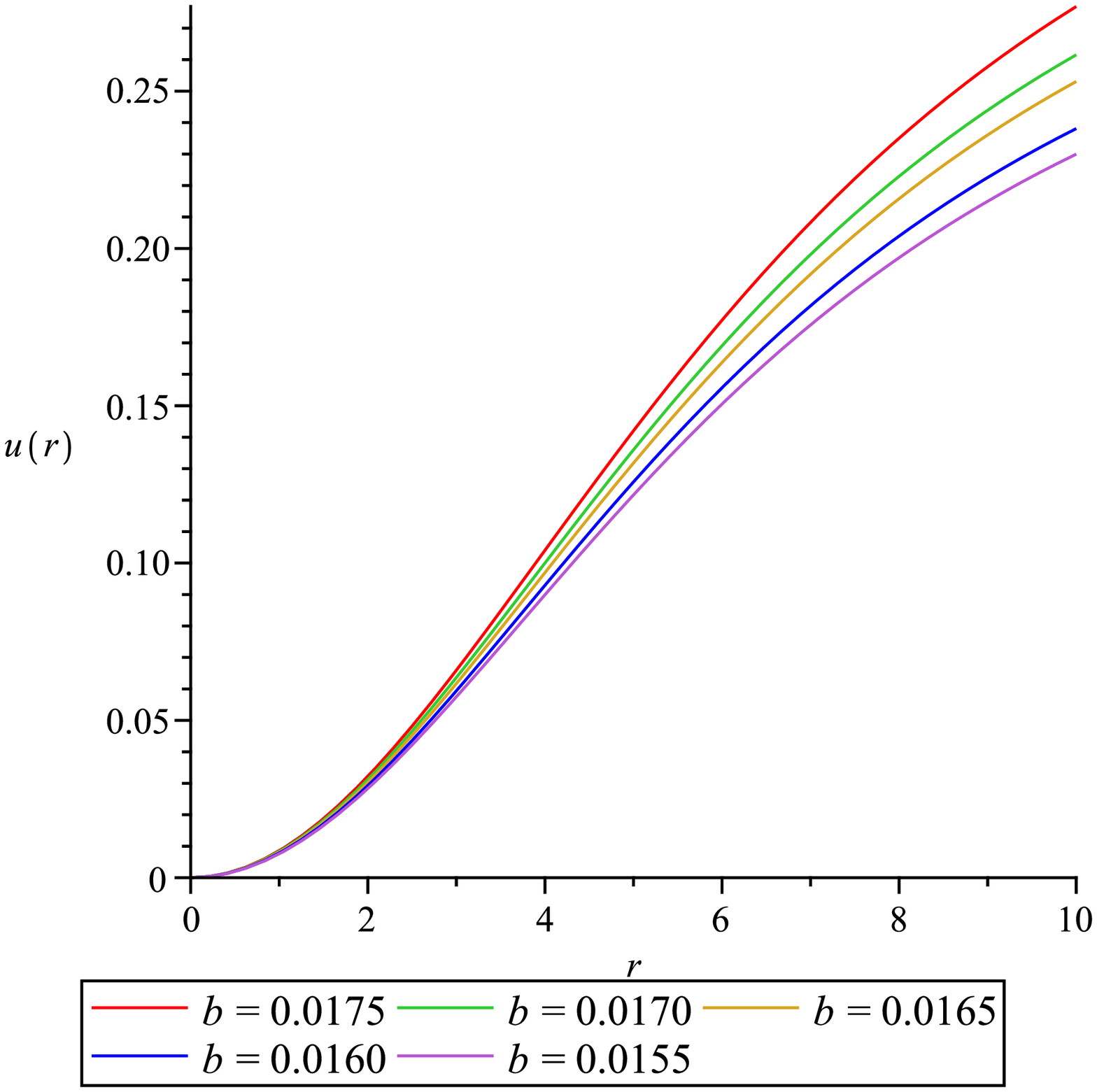}
\includegraphics[width=5.0cm]{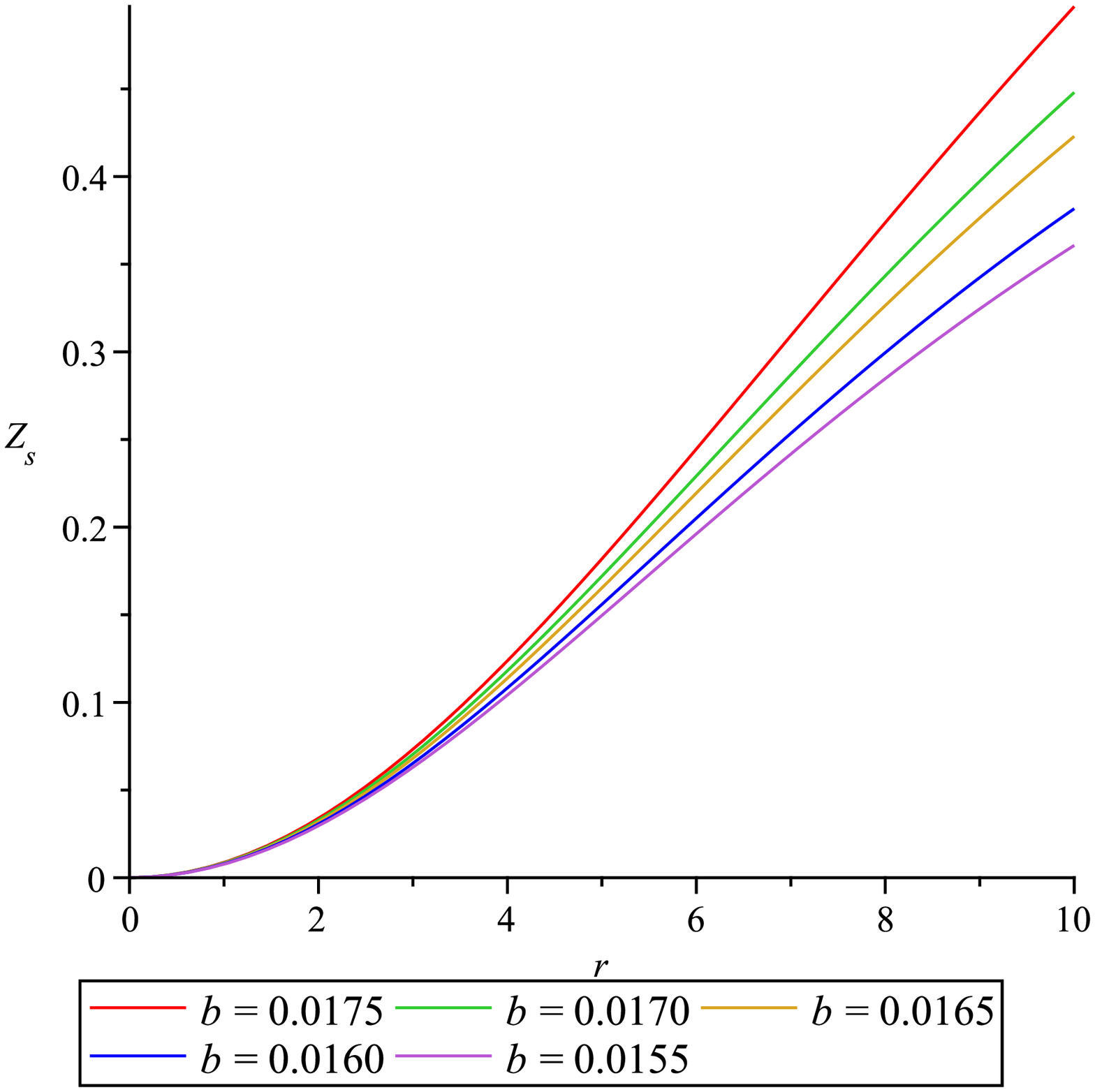}\\
\end{tabular}
\caption{ (Left) Variation of mass vs radial coordinate is shown
in the plot for the specified range for positive flag curvature.
(Middle) Variation of the compactness vs radial coordinate is
shown in the plot for the specified range for positive flag
curvature. (Right) Variation of the redshift vs radial coordinate
is shown in the plot for the specified range for positive flag
curvature. }
\end{figure*}

\begin{table}
\caption{Maximum compactness factor, mass and surface redshift for
different cases.  } \label{tab:3}
\begin{tabular}{@{}llllll@{}}
\hline ~~~~a   &  ~~~~ b  & ~~ u(R) & ~~m(R)  & ~~~~Z(R)\\ & & &
(in km) & \\ \hline
 0.0216 & ~~0.0175 &~~ 0.2769 & ~~2.769 & ~~0.4970 \\
  0.0225 & ~~0.0170 & ~~0.2615 & ~~2.615 & ~~0.4480 \\
  0.0226 &~~ 0.0165 &~~ 0.2531 & ~~2.531 & ~~0.4229  \\
  0.0236 &~~ 0.0160 &~~ 0.2381 &~~ 2.381 & ~~0.3817 \\
  0.0237 &~~ 0.0155 &~~ 0.2299  &~~ 2.299  & ~~0.3607 \\
 \hline
\end{tabular}
\end{table}

\subsection{Energy Condition}

Now, we verify the energy conditions namely null energy condition
(NEC), weak energy condition (WEC) and strong energy condition
(SEC) which can be given as follows:
\begin{equation}
(i)~NEC:\rho+p_r\geq 0,
\end{equation}

\begin{equation}
(ii)~WEC:\rho+p_r\geq0,~~\rho\geq 0,
\end{equation}

\begin{equation}
(iii)~SEC: \rho+p_r\geq0,~~~ \rho+p_r+2p_t\geq 0.
\end{equation}

We plot the L.H.S of the above inequalities in Fig. 3 which shows
that these inequalities hold good. This therefore confirm that our
model satisfies all the energy conditions.

\begin{figure*}[thbp]
\begin{tabular}{rl}
\includegraphics[width=4.0cm]{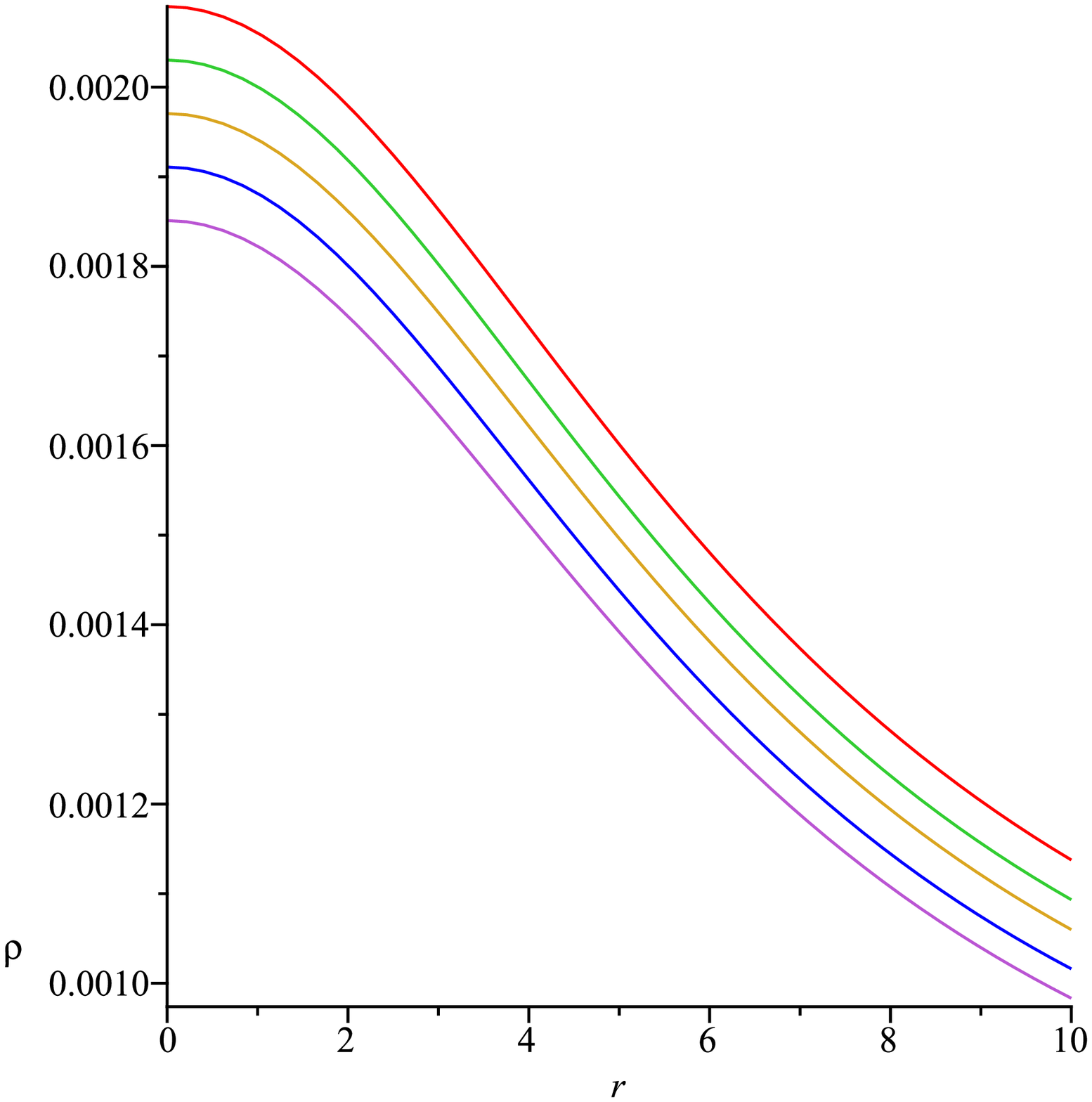}&
\includegraphics[width=4.0cm]{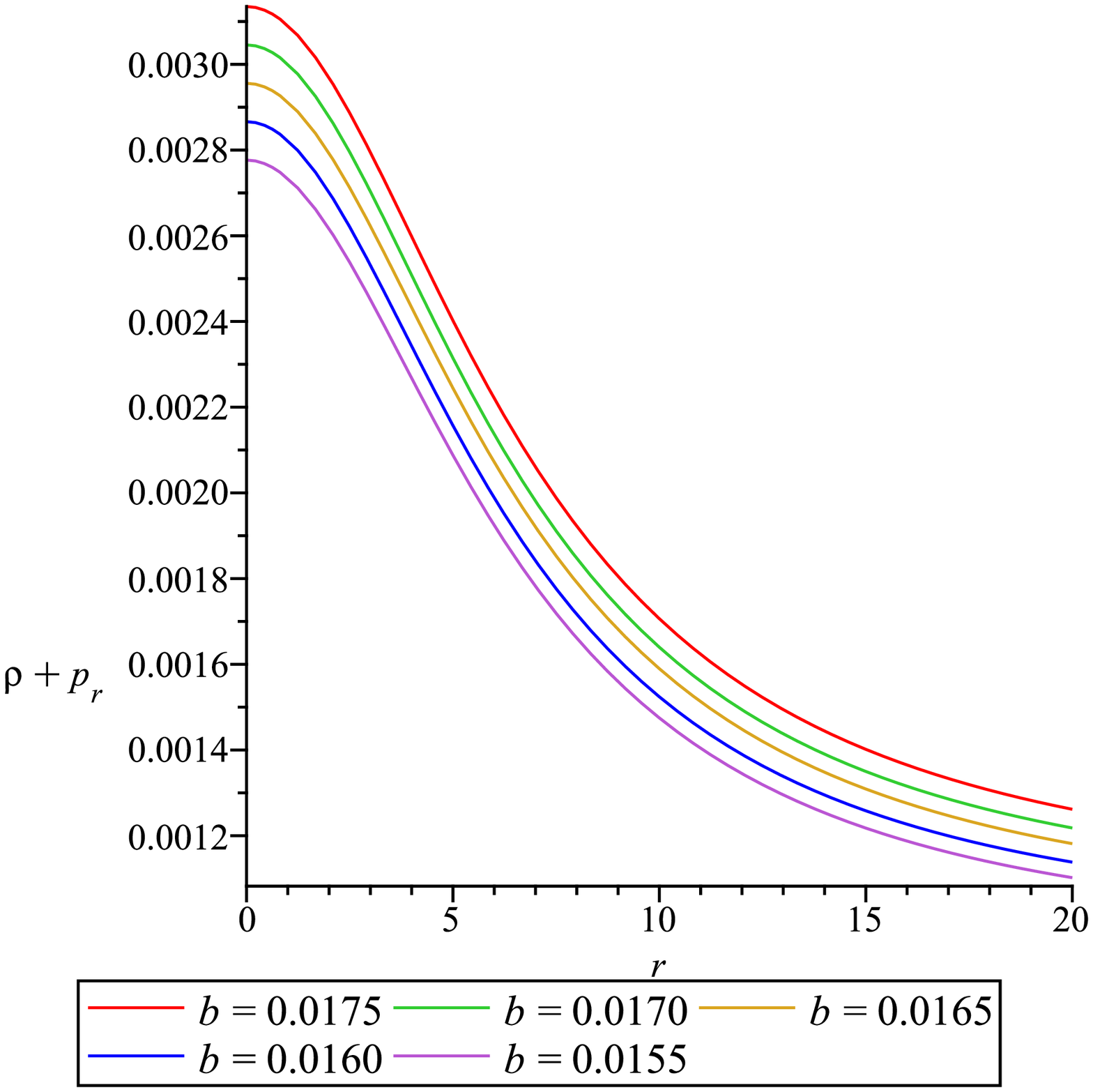}
\includegraphics[width=4.0cm]{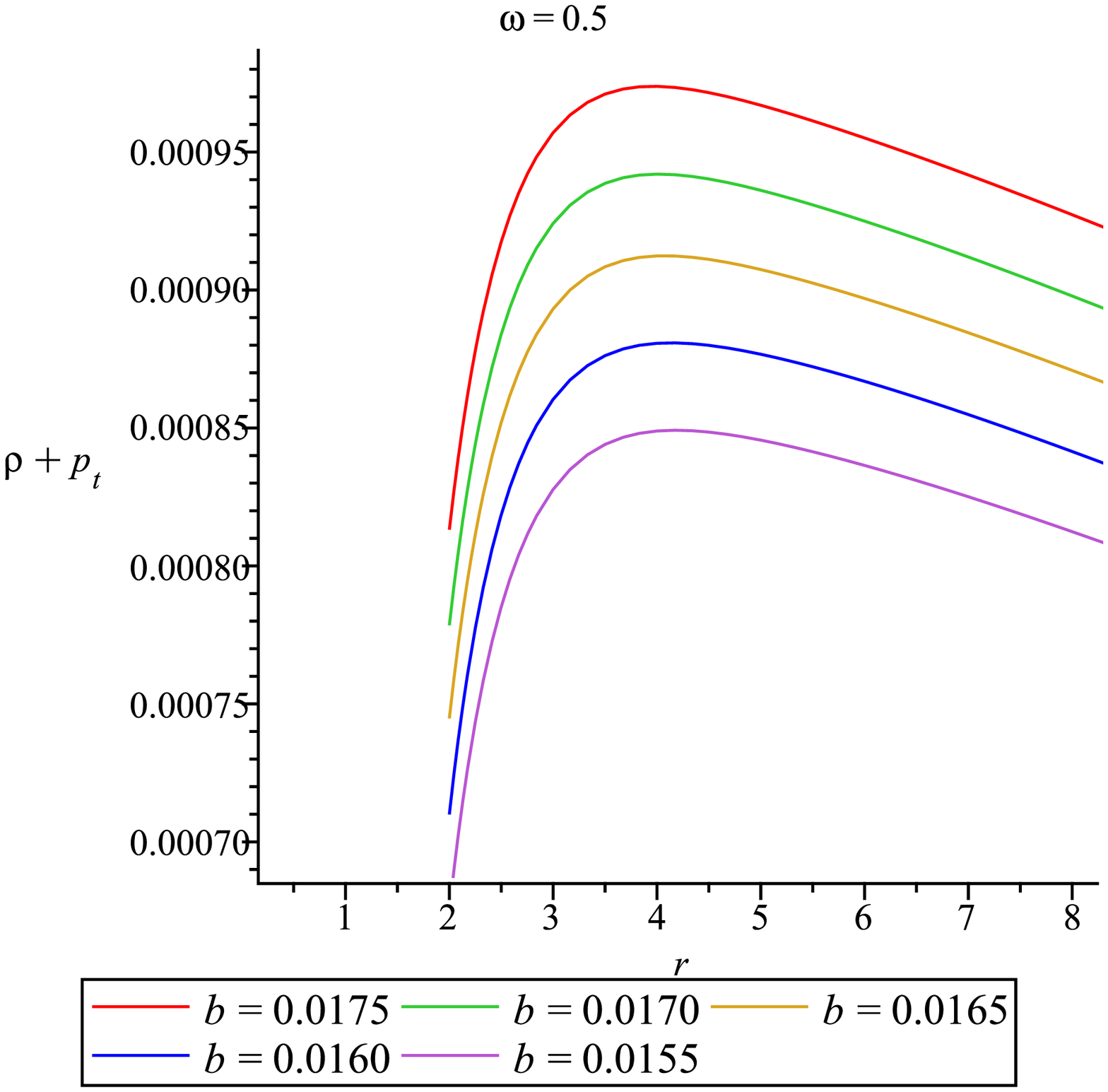}
\includegraphics[width=4.0cm]{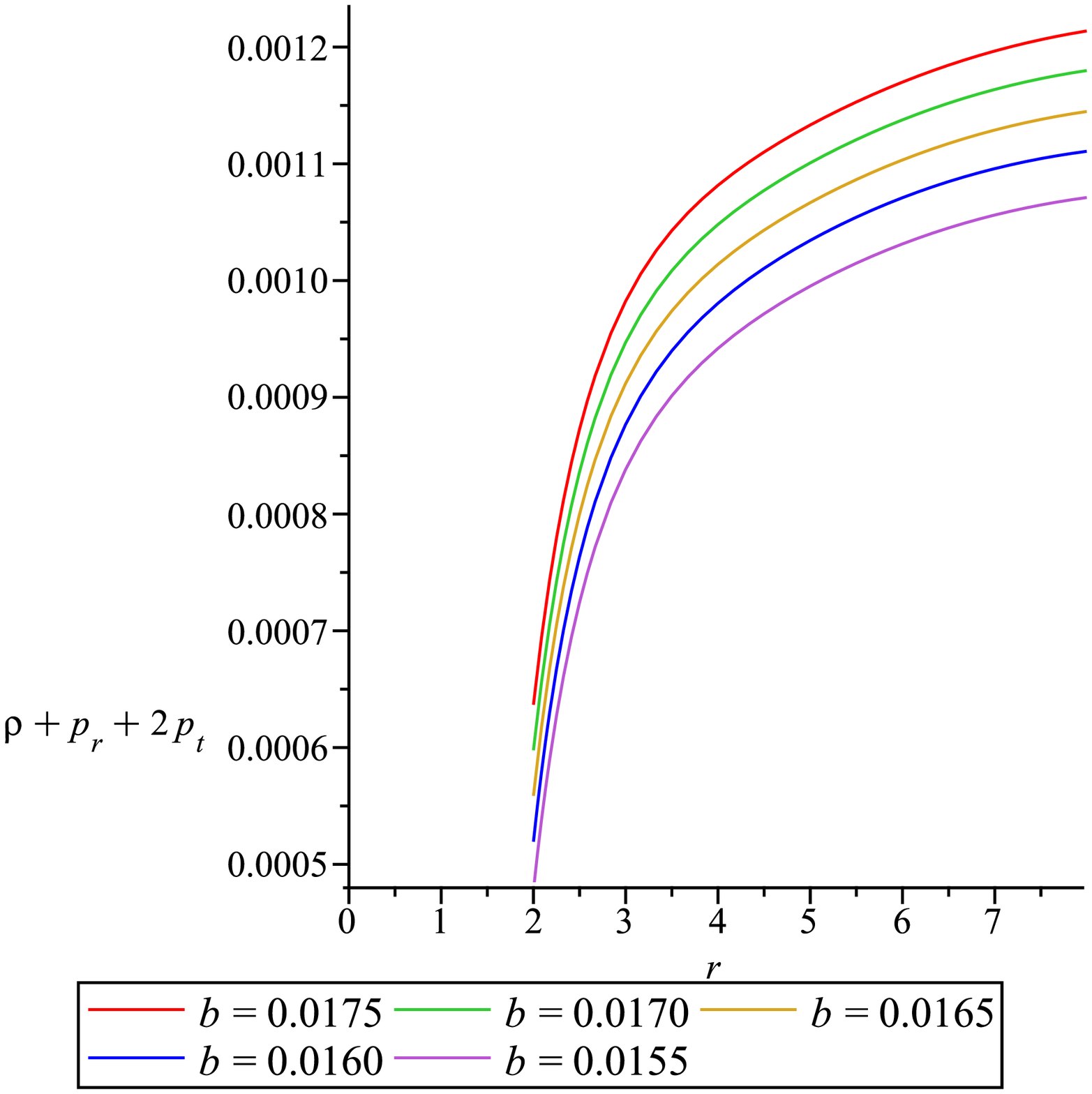}\\
\end{tabular}
\caption{ (Left) Variation of $\rho $ vs radial coordinate is
shown in the plot for the specified range for positive flag
curvature.(First Middle) Variation of $\rho+p_r$ vs radial
coordinate is shown in the plot for the specified range for
positive flag curvature. (Second  Middle) Variation of $\rho+p_t$
vs radial coordinate is shown in the plot for the specified range
for positive flag curvature. (Right) Variation of $\rho+p_r+2p_t$
vs radial coordinate is shown in the plot for the specified range
for positive flag curvature. }
\end{figure*}

\subsection{TOV Equation}

The generalized Tolman-Oppenheimer-Volkoff (TOV) equation for this
system can be given by the equation \cite{Varela2010}
\begin{equation}
-\frac{M_G(\rho+p_r)}{r^{2}} \sqrt{\frac{A}{B}}-\frac{dp_r}{dr}+\frac{2}{r}(p_t-p_r)=0,
\end{equation}
where $M_G=M_G(r)$ is the effective gravitational mass inside a
sphere of radius $r$ given by the Tolman-Whittaker formula which
can be derived from the equation
\begin{equation}
M_G(r)=\frac{1}{2}r^{2} \frac{B'}{\sqrt{AB}}.
\end{equation}
The above equation explains the equilibrium condition of the fluid
sphere due to combined effect of gravitational, hydrostatics and
anisotropy forces. Equation $(34)$ can be rewritten in the
following form
\begin{equation}
F_g+F_h+F_a=0,
\end{equation}
where
\begin{equation}
F_g=-\frac{B'}{2B}(\rho+p_r),
\end{equation}

\begin{equation}
F_h=-\frac{dp_r}{dr},
\end{equation}

\begin{equation}
F_a=\frac{2}{r}(p_t-p_r).
\end{equation}

The profiles (Fig. 3) of these force indicate that the matter
distribution comprising the compact star is in equilibrium state
subject to the gravitational force $F_g$, hydrostatic force $F_h$
plus another force $F_a$ due to anisotropic pressure. The first
two forces are repulsive in nature due to positivity but the later
force is in attractive nature. The combined effect of these forces
make the system in a equilibrium position.

 \begin{figure*}[thbp]
\begin{tabular}{rl}
\includegraphics[width=5.0cm]{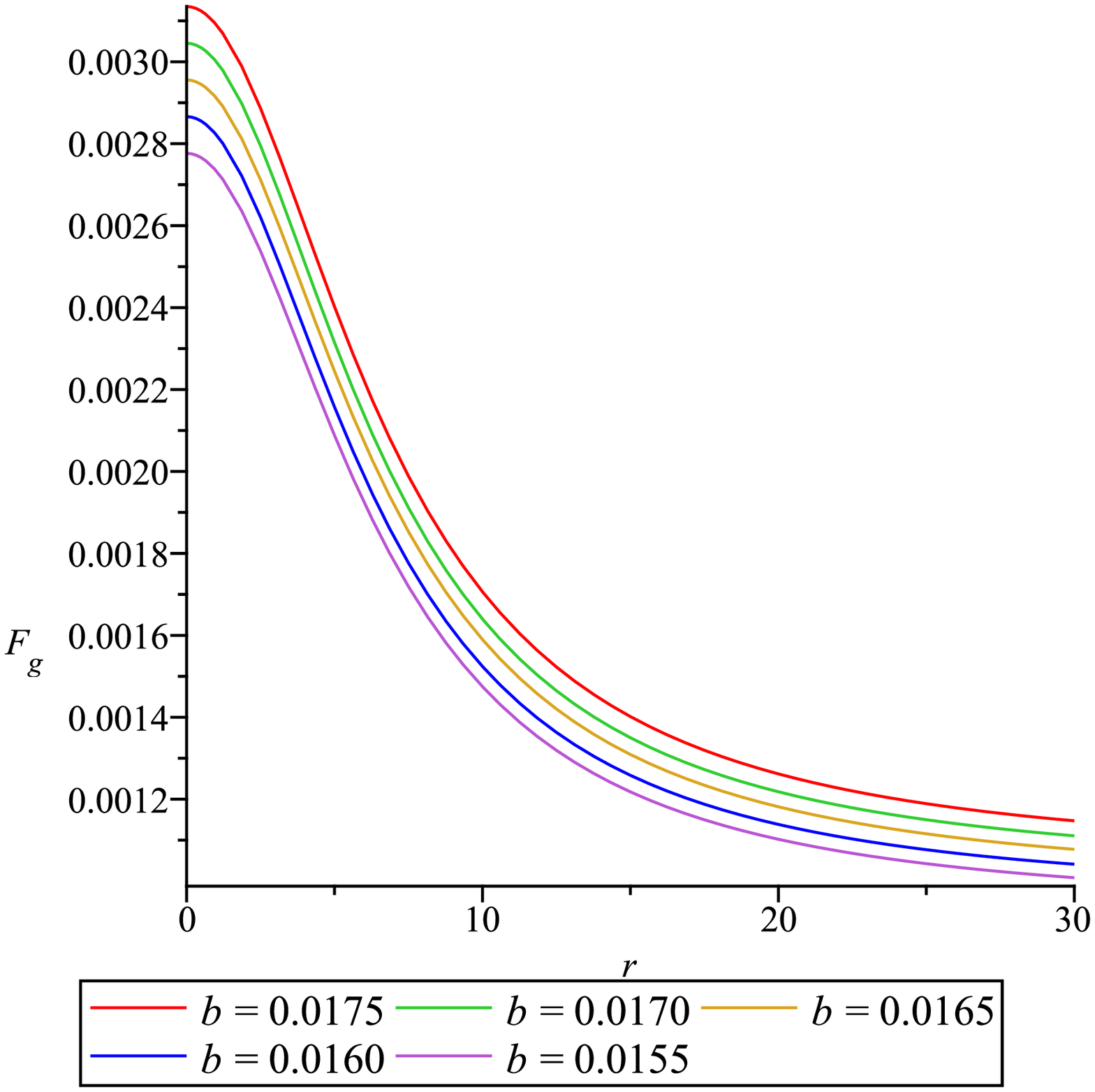}&
\includegraphics[width=5.0cm]{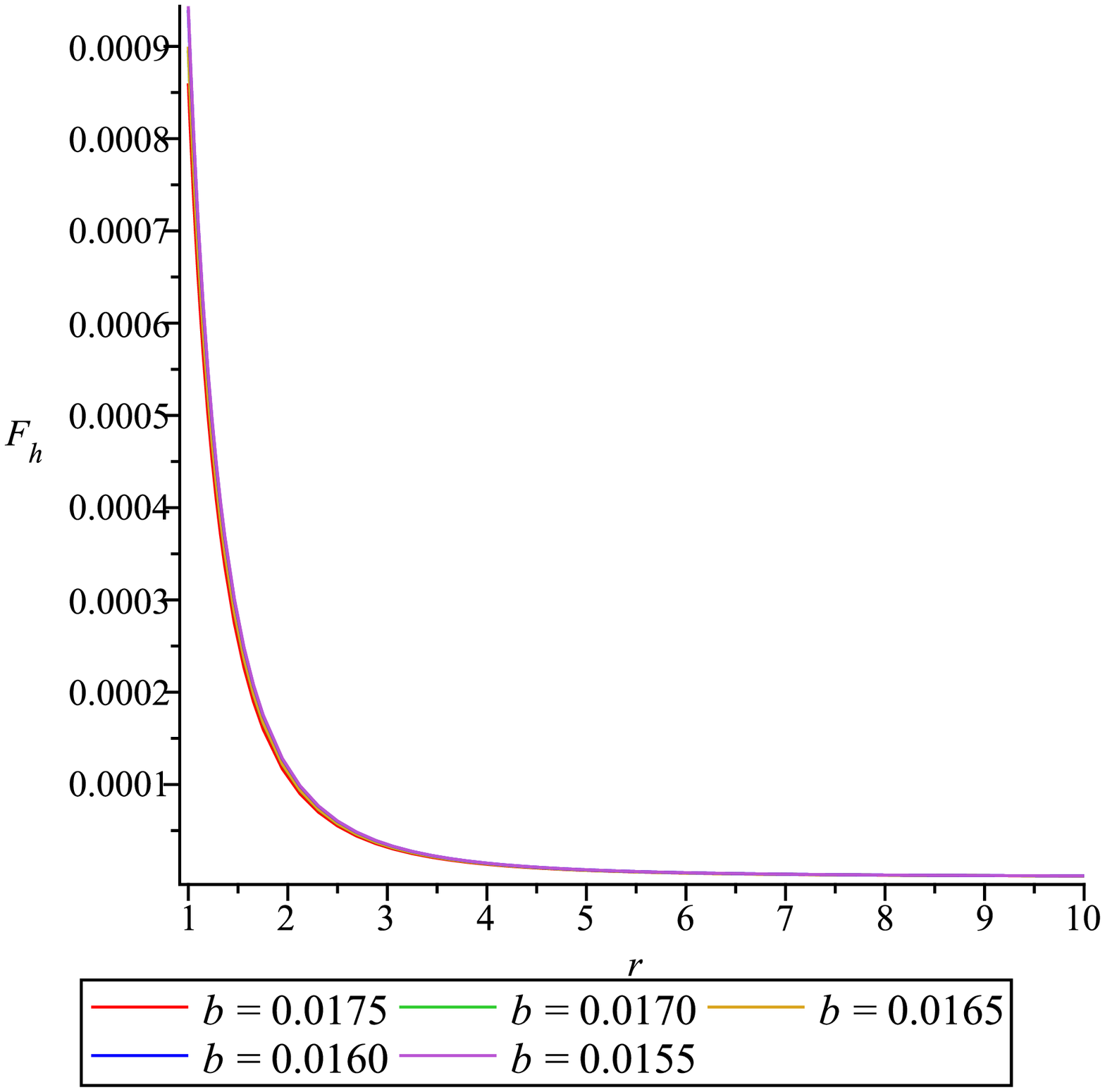}
\includegraphics[width=5.0cm]{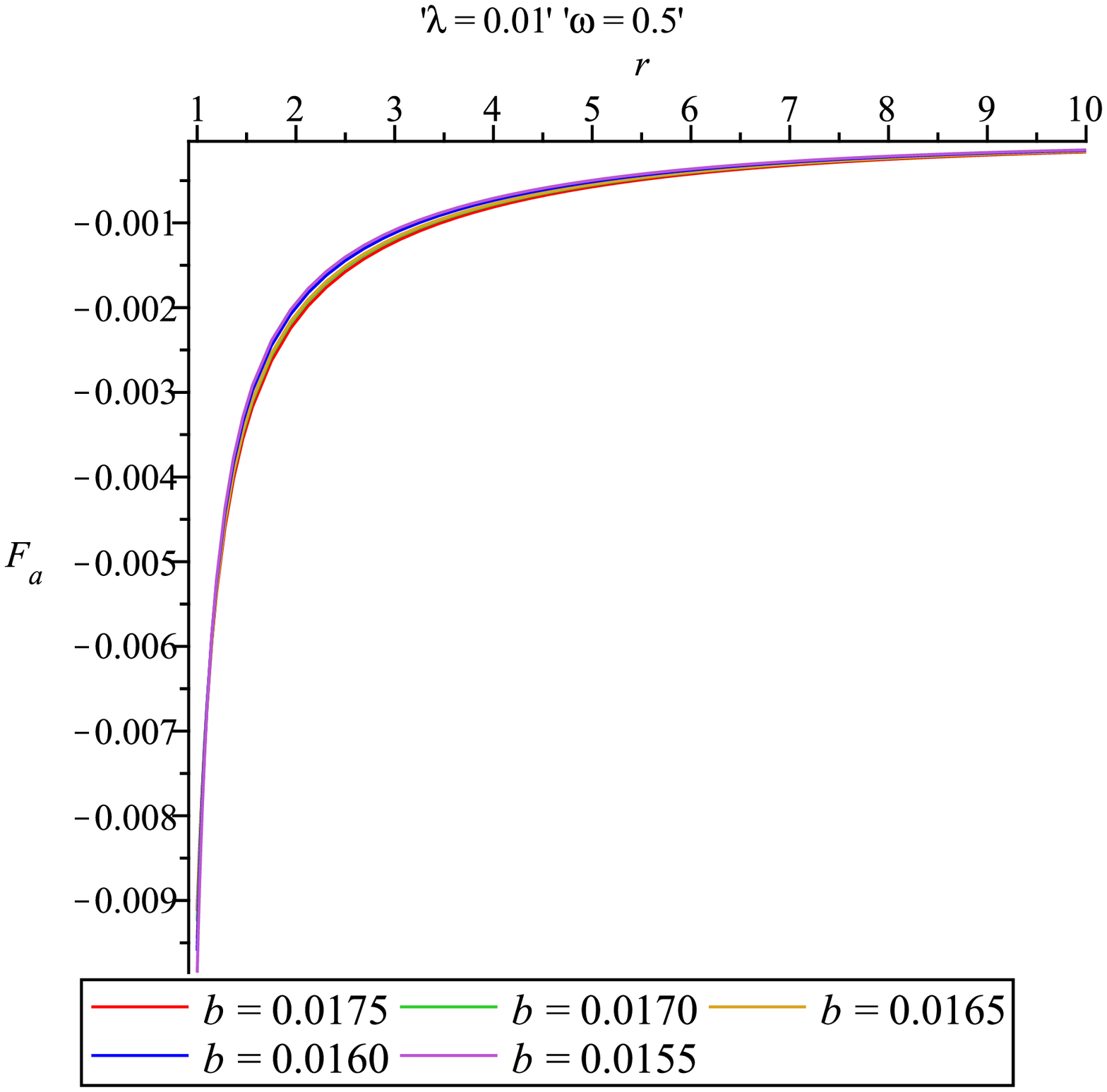}\\
\end{tabular}
\caption{ (Left) Variation of the gravitational force vs radial
coordinate is shown in the plot for the specified range for
positive flag curvature. (Middle) Variation of hydrostatics force
vs radial coordinate is shown in the plot for the specified range
for positive flag curvature. (Right) Variation of anisotropic
force vs radial coordinate is shown in the plot for the specified
range for positive flag curvature. }
\end{figure*}

\subsection{Stability}

Now, we  examine the stability of model. For this purpose,  we
employ the technique proposed by Herrera \cite{Herrera1992} which
is known as cracking concept. At first, it requires that the
squares of the radial and tangential sound speeds should be within
the limit $[0,1]$. The theorem states that one can get stable
configuration if radial speed of sound is greater than that of
transverse speed, i.e. $v_{st}^2 - v_{sr}^2$ should less than zero
within the matter distribution.

Now, we calculate the radial speed ($v_{sr}$)  and transverse
speed ($v_{st}$) for  our anisotropic model as
\begin{equation}
v_{sr}^{2}=\frac{dp_r}{d\rho}=\omega,
\end{equation}

\begin{equation}
v^2_{st}=\frac{dp_t}{d\rho} =\frac{\alpha+\beta+\gamma}{2abr(5+ar^2)},
\end{equation}
where $\alpha = \frac{16ab\omega\lambda r-16\omega^2a^2b\lambda
r^3+56Gwab^2r^3+6Gb^2r+16Gab^2r^3+6Ga^2b^2r^5+18G\omega^2b^2r+24G\omega^2ab^2r^3+6G\omega^2a^2b^2r^5+12G\omega
a^2b^2r^5}{4(\lambda+a\lambda r^2-Gbr^2)}$,

$ \beta = \frac{6ar(12\omega b\lambda+8ab\omega\lambda
r^2-4\omega^2a^2b\lambda
r^4+14Gwab^2r^4+3Gb^2r^2+4Gab^2r^4+Ga^2b^2r^6+9G\omega^2b^2r^2+6G\omega^2ab^2r^4+G\omega^2a^2b^2r^6+2G\omega
a^2b^2r^6)}{4(1+ar^2)(\lambda+a\lambda r^2-Gbr^2)}$,

$ \gamma = \frac{(2a\lambda r-2Gbr)(12\omega
b\lambda+8ab\omega\lambda r^2-4\omega^2a^2b\lambda
r^4+14Gwab^2r^4+3Gb^2r^2+4Gab^2r^4+Ga^2b^2r^6+9G\omega^2b^2r^2+6G\omega^2ab^2r^4+G\omega^2a^2b^2r^6+2G\omega
a^2b^2r^6)}{4(\lambda+a\lambda r^2-Gbr^2)^2}$.

To check whether the sound speeds lie between 0 and 1 and
$v_{st}^2 - v_{sr}^2 <0$ we plot the radial and transverse sound
speeds and squares of there difference. Fig. 4 satisfies Herrera's
criterion and therefore, our model is quite stable one.

\begin{figure*}[thbp]
\begin{tabular}{rl}
\includegraphics[width=5.0cm]{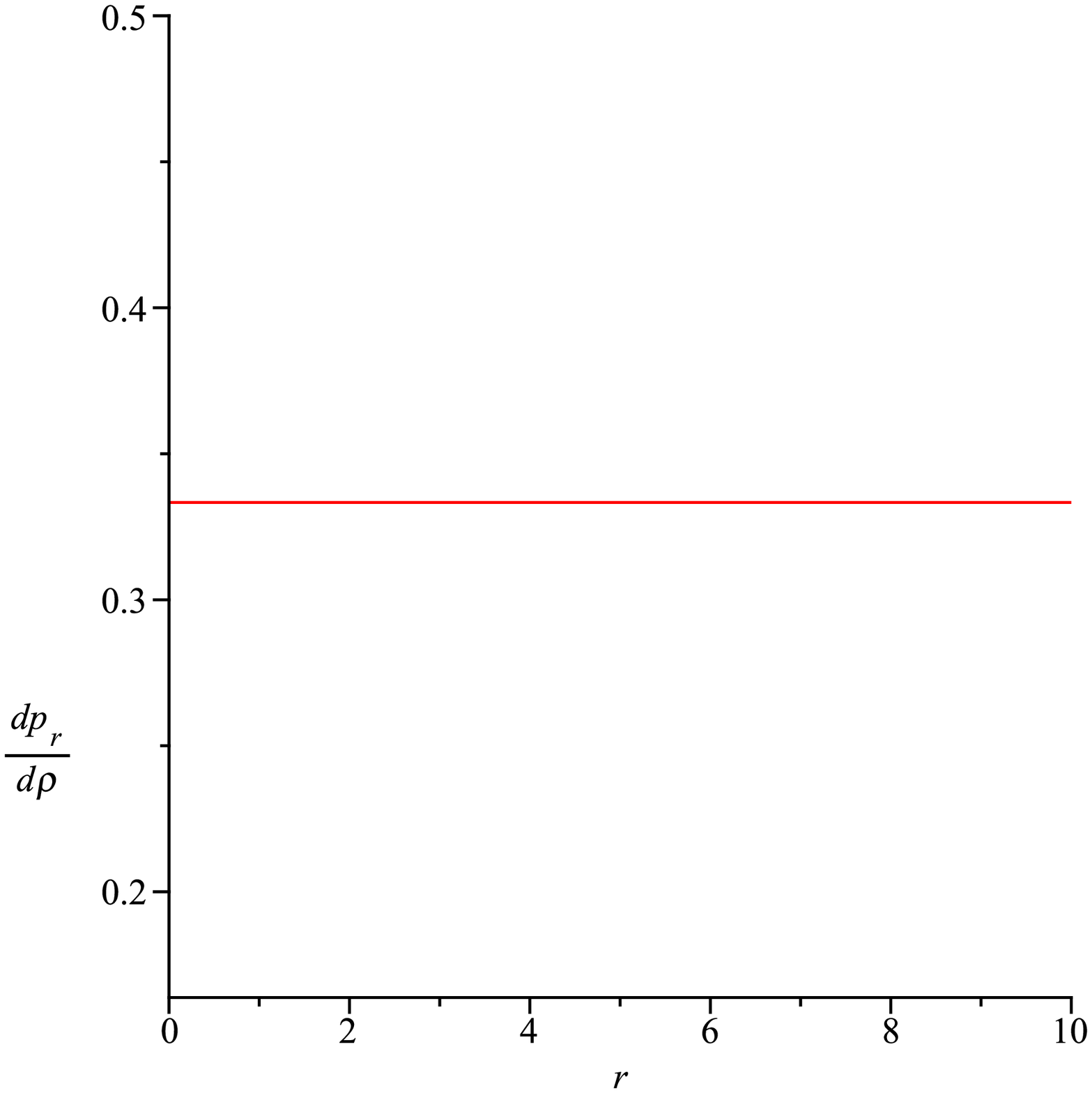}&
\includegraphics[width=5.0cm]{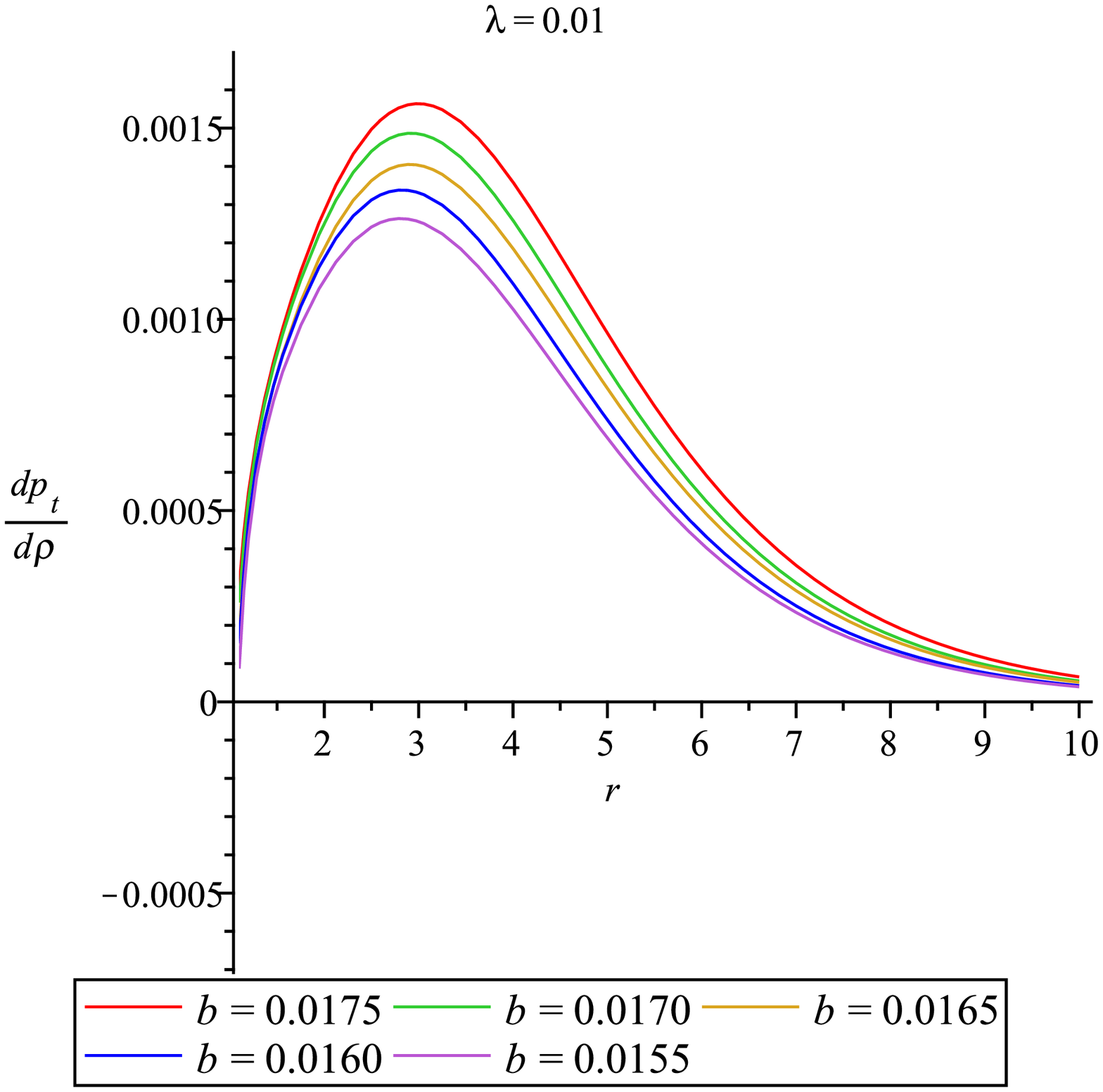}
\includegraphics[width=5.5 cm]{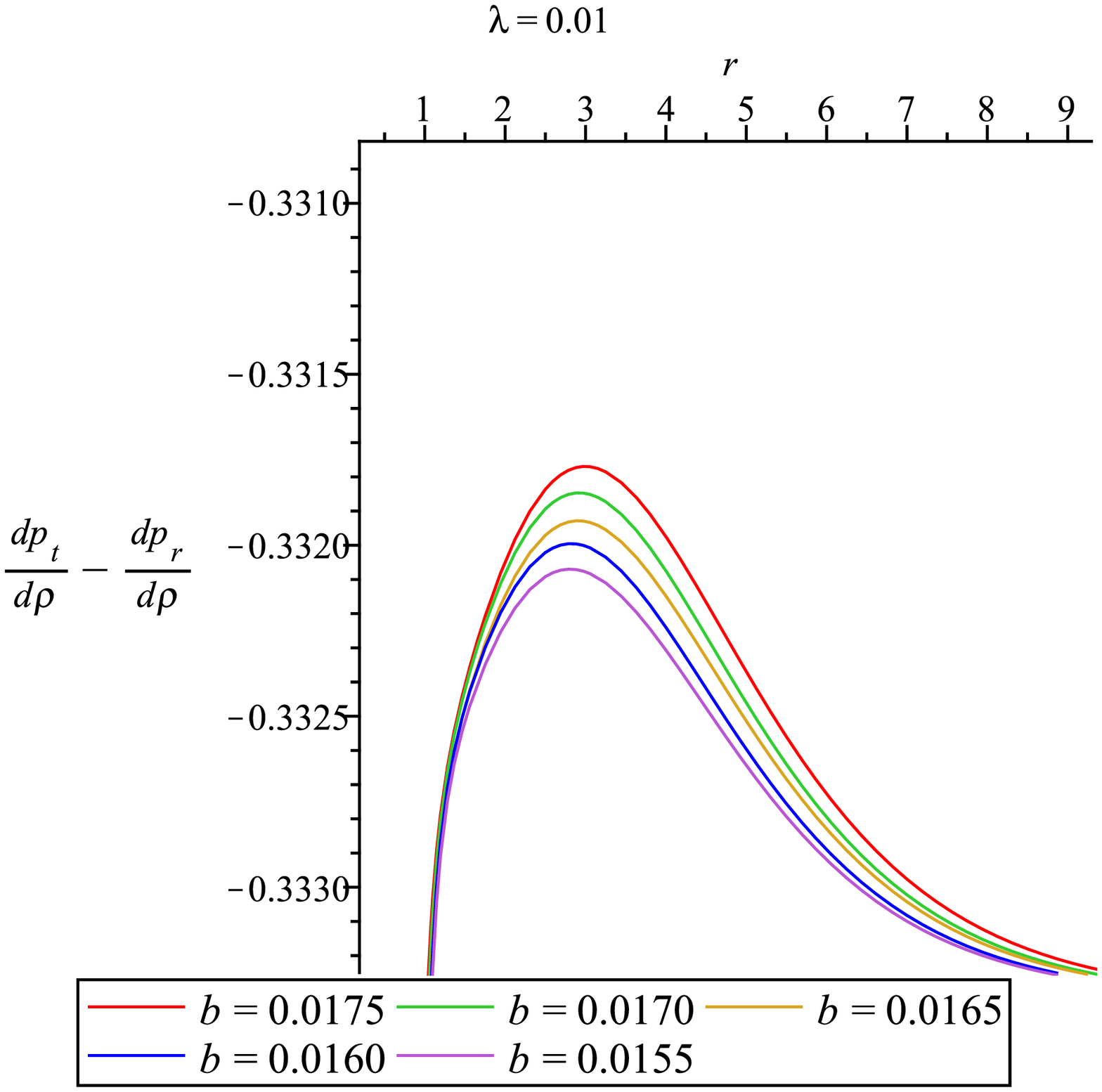}\\
\end{tabular}
\caption{ (Left) Variation of the square of the radial velocity vs
radial coordinate is shown in the plot for the specified range for
positive flag curvature. (Middle) the square of the transverse
velocity. (Right) Variation of  the difference between the square
of the transverse velocity and square of the radial velocity vs
radial coordinate is shown in the plot for the specified range for
positive flag curvature. }
\end{figure*}

\section{CONCLUDING REMARKS}

In the present investigation, we have considered anisotropic matter
source for constructing a new type of solutions for compact stars.
The background geometry is taken as the Finslerian structure of
spacetime. It is expected that the compactness of these stars is
greater than that of a neutron star. Plugging the expressions for
$G$ and $c$ in the relevant equations, one can figure out that the
value of the central density for the choices of the constant $b$
turn out to be $\rho_0 \varpropto ~ 10^{15}$ gm~cm$^{-3}$ which is
in observation relevance
\cite{Ruderman1972,Glendenning1997,Herjog2011}. This result is
hopeful as far as physical aspect is concerned and may be treated as
a seminal bottom line of the present study.

In this same physical point of view we have studied several other
physical aspects of the model to justify validity of the solutions.
The features that emerge from the present investigation can be
put forward as follows:

(1) Mass-radius relation: The surface redshift, which gives the
compactness of the star, comes out to be in the range $0.36-0.49$
in our study. This high redshift are consistent with strange stars
which have mass-radius ratios higher than neutron stars
\cite{Lindblom1984}.

In this connection we were also curious about the condition of
Buchdahl \cite{Buchdahl1959} related to maximum allowable
mass-radius ratio limit. It is observed that Buchdahl's limit
has been satisfied by our model.

(2) Energy Condition: In the present model satisfies all the
energy conditions are shown to be satisfactory.

(3) TOV Equation: The generalized Tolman-Oppenheimer-Volkoff equation
for the Finslerian system of compact star are studied. It has been observed
that by the combined effect of the forces in action keep the system
in static equilibrium.

(4) Stability: The stability of model has been examined by
employing the cracking technique of Herrera \cite{Herrera1992}.
We have shown via Fig. 4 that Herrera's criterion satisfies
which therefore indicates stability of our model.

As a final comment we would like to mention that the toy model as
put forward in the present study for compact stars under the
Finslerian structure of spacetime are seem very promising.
However, some other aspects are deemed to be performed, such as
issues of formation and structure of various compact stars, before
one could be confirmed about the satisfactory role of the
Finslerian spacetime than that of Riemannian geometry.
Specifically several other issues as argued by Pfeifer and
Wohlfarth \cite{Pfeifer2012} ``Finsler spacetimes are viable
non-metric geometric backgrounds for physics; they guarantee well
defined causality, the propagation of light on a non-trivial null
structure, a clear notion of physical observers and the existence
of physical field theories determining the geometry of space-time
dynamically in terms of an extended gravitational field equation''
can be sought for in a future study.

\subsection*{Appendix}

Let us choose $\bar{F^2}$ in the following form

 \[  \bar{F^2}=y^\theta y^\theta+f(\theta,\phi)y^\phi y^\phi
,\] That is, \[ \bar{g}_{ij}=diag(1,f(\theta,\phi)),~~~~~~~and~~~~~ \bar{g}^{ij}=diag(1,\frac{1}{f(\theta,\phi)});~~~~~(i,j= 2,3 /\theta,
\phi)\]
One can find from $\bar{F^2}$
\[ G^2=-\frac{1}{4}\frac{\partial f}{\partial \theta}\]

\[ G^3=\frac{1}{4f}\left(2\frac{\partial f}{\partial
\theta}y^\phi y^\theta+\frac{\partial f}{\partial
\phi}y^\phi y^\phi\right)\]
Hence, one obtains
\[ \bar{F}^2\bar{Ric}=y^\phi y^\phi \left[-\frac{1}{2}\frac{\partial^2
f}{\partial \theta^2}  +\frac{1}{2f}\frac{\partial^2 f}{\partial
\phi^2}-\frac{1}{2}\frac{\partial}{\partial
\phi}\left(\frac{1}{f} \frac{\partial f}{\partial
\phi}\right) - \frac{1}{4f} \left(\frac{\partial f}{\partial
\theta}\right)^2+\frac{1}{4f}\frac{\partial f}{\partial
\phi}\frac{1}{f}\frac{\partial f}{\partial \phi}+\frac{\partial
f}{\partial \theta}\frac{1}{2f}\frac{\partial f}{\partial
\theta}-\frac{1}{4f^2}\left(\frac{\partial f}{\partial
\phi}\right)^2\right]\]

 \[+y^\theta
y^\theta\left[-\frac{1}{2}\frac{\partial}{\partial
\theta}\left(\frac{1}{f} \frac{\partial f}{\partial
\theta}\right)-\frac{1}{4f^2}\left(\frac{\partial f}{\partial
\theta}\right)^2\right]+ y^\phi y^\theta\left[\frac{1}{f}\frac{\partial^2f}{\partial \theta \partial
\phi}-\frac{1}{2}\frac{\partial}{\partial
\theta}\left(\frac{1}{f} \frac{\partial f}{\partial
\phi}\right)-\frac{1}{2}\frac{\partial}{\partial
\phi}\left(\frac{1}{f} \frac{\partial f}{\partial
\theta}\right)\right]\]

  Now, coefficient of  $y^\phi y^\theta =0$ iff, $ f$  is independent of $\phi$  i.e.
  \[f(\theta, \phi) =f(\theta)  \]
( but coefficient of $y^\theta y^\theta $ $\&$  $y^\phi y^\phi $ are non zero )

Therefore,
\[ \bar{F}^2\bar{Ric}=\left[-\frac{1}{2f}\frac{\partial^2
f}{\partial \theta^2}  +\frac{1}{4f^2}\left(\frac{\partial f}{\partial
\theta}\right)^2\right] (y^\theta y^\theta+f y^\phi y^\phi)\]
Hence,
\[  \bar{Ric}= -\frac{1}{2f}\frac{\partial^2
f}{\partial \theta^2}  +\frac{1}{4f^2}\left(\frac{\partial f}{\partial
\theta}\right)^2  = \lambda  \]
Here, $\lambda$ may be a constant or a function of $\theta$.

Putting, $\frac{1}{f}\left(\frac{\partial f}{\partial
\theta}\right) =T(\theta)$, the above equation yields
\[\frac{dT}{d \theta}+\frac{1}{2}T^2+2 \lambda =0\]

For constant $\lambda$, one can get Finsler structure $\bar{F^2}$ as
\[\bar{F}^2 = y^\theta y^\theta  + A \sin^2(\sqrt{\lambda} \theta )y^\phi y^\phi,~~\lambda > 0 \]
\[ = y^\theta y^\theta  + A \theta^2 y^\phi y^\phi,~~\lambda = 0 \]
\[= y^\theta y^\theta  + A \sinh^2(\sqrt{-\lambda} \theta )y^\phi y^\phi,~~\lambda < 0 \]
[ A may be taken as 1 ]\\

Now, the Finsler structure takes the form
\[ {F}^2 =B(r) y^ty^t - A(r) y^ry^r -r^2 y^\theta y^\theta  - r^2\sin^2 \theta y^\phi y^\phi  + r^2\sin^2 \theta y^\phi y^\phi -r^2\sin^2(\sqrt{\lambda} \theta )y^\phi y^\phi\]
\[= \alpha^2 + r^2 (\sin^2 \theta - \sin^2(\sqrt{\lambda} \theta ))y^\phi y^\phi~~~~~~~~~~~~~~~~~~~~~~~~~~~~~~~~~~~~~~~~~~~~~~~~\]
\[= \alpha^2 + r^2 \chi (\theta)y^\phi y^\phi~~~~~~~~~~~~~~~~~~~~~~~~~~~~~~~~~~~~~~~~~~~~~~~~~~~~~~~~~~~~~~~~~~~ \]
where  $\chi (\theta) = \sin^2 \theta - \sin^2(\sqrt{\lambda} \theta )$.

Thus,
\[ {F} = \alpha \sqrt{ 1 + \frac{r^2 \chi (\theta)y^\phi y^\phi}{\alpha^2}}\]
Let, $b_\phi= r\sqrt{ \chi (\theta)} $, then
\[ {F} = \alpha \sqrt{ 1 + \frac{(b_\phi y^\phi)^2}{\alpha^2}} = \alpha \sqrt{ 1 +s^2} \]
where, \[  s =\frac{(b_\phi y^\phi)}{\alpha} = \frac{\beta}{\alpha} \]
\[ b_\mu = ( 0,0,0,b_\phi)  ~,~ ~ b_\phi y^\phi = b_\mu h^\mu = \beta  ~, ~ (\beta ~is ~one~ form ) \]
Finally, we have
\[F = \alpha \phi(s)~~ , ~~\phi(s) = \sqrt{1+s^2}\]

Hence, F is the metric of $(\alpha, \beta)$-Finsler space.
\\

   The killing equation $K_V(F) =0 $ in   Finsler space can be obtained by considering the isometric transformations of Finsler structure \cite{XL}. One can investigate the Killing vectors of $(\alpha, \beta)$ - Finsler space. The Killing equations for this class of Finsler space is given as
    \[\left(\phi(s)-s\frac{\partial \phi(s)}{\partial s}\right)K_V(\alpha) + \frac{\partial \phi(s)}{\partial s }K_V\beta) =0, \]
   where
   \[K_V(\alpha) = \frac{1}{2 \alpha}\left(V_{\mu \mid \nu} +V_{\nu \mid \mu} \right)y^\mu y^\nu,\]
    \[K_V(\beta) =  \left(V^\mu \frac{\partial b_\nu}{\partial x^\mu } +b_{ \mu}\frac{\partial V^\mu}{\partial x^\nu} \right)y^\nu.\]
   Here $"\mid"$ represents the covariant derivative with respect to the Riemannian metric $\alpha$. For the present case of Finsler structure it is given by
   \[K_V(\alpha)+ sK_V(\beta)=0 ~~or~~  \alpha K_V(\alpha)+  \beta K_V(\beta)=0.\]
   Consequently, we have the solution
   \[  K_V(\alpha)= 0 ~~and~~K_V(\beta)=0\]
   or
   \[ V_{\mu \mid \nu} +V_{\nu \mid \mu} =0\]
   and
   \[ V^\mu \frac{\partial b_\nu}{\partial x^\mu } +b_{ \mu}\frac{\partial V^\mu}{\partial x^\nu} =0.\]
   It is to be noted that the second Killing equation constrains the first one which is, in fact, the Killing equation of the Riemannian space, that is, it is responsible for breaking the symmetry (isometric) of the Riemannian space.

   On the othe hand, the Finsler space we are considering is , in fact, can be determined from a Riemannian manifold  $( M, g_{\mu \nu}(x))$ as we have
   \[F(x,y) =\sqrt{g_{\mu \nu}(x)y^\mu y^\nu }\]
( cf. equations (5) and (6) in the case $\bar{F^2}$ is quadric in $ y^\theta~ \& ~y^\phi $    )
 \\

 It is a semi-definite Finsler space. Therefore, we can take covariant
derivative of the Riemanian space. The Bianchi identities are, in
fact, coincident with those of the Riemanian space (being the
covariant conservation of Einstein tensor). The present Finsler space
is reducible to the Riemanian space and consequently the gravitational
field equations can be obtained. Also  we shall find the gravitational field
equations alternatively
following \cite{XC}. They have also shown
the covariantly conserved properties of the tensor $G^\mu_\nu$  in respect
of covariant derivative in Finsler spacetime with the Chern-Rund
connection. Presently this conserved property of $G^\mu_\nu$ which are, in
fact, in the same forms but obtained from the Riemanian manifold
follows by using the covariant derivative of that space (which are, in
fact, the Bianchi identity). Also we point out the gravitational field
equation (18) is restricted to the base manifold of the Finsler space,
as in \cite{XL}, and the fiber coordinates $y^i$ are set to be
the velocities of the cosmic components (velocities in the energy
momentum tensor). Also, Xin  Li, et al. \cite{XL} have shown that
there gravitational field equation could be derived from the of
Pfeifer et.al. approximately \cite{Pf}. Pfeifer et.al. \cite{Pf} have constructed
gravitational dynamic for Finsler spacetime in terms of an action
integral on the unit tangent bundle. Also the gravitational field equation (18) is insensitive to the connection because $G_\nu^\mu$ are obtained from the Ricci scalar which is in fact, insensitive to the connections and depend only on the Finsler structure F.

Thus the above equations(20)-(22) are derived from the modified gravitational field equation (18) taking anisotropic energy momentum tensor (19) as well as these equations are derivable from the Einstein gravitational field equation in the Riemannian spacetime with the metric (6) in which the metric $\bar{g}_{ij}$ is given by
\[\bar{g}_{ij} = diag~(~ 1~,~~ \sin^2 \sqrt{\lambda} \theta~ )\]

   That is,
   \[g_{\mu\nu }= diag~(~B,~-A,~ -r^2~,~~ -r^2\sin^2 \sqrt{\lambda} \theta~ )\]
   \[g^{\mu\nu }= diag~(~B^{-1},~-A^{-1},~ -r^{-2}~,~~ -r^{-2}\sin^{-2} \sqrt{\lambda} \theta~ )\]

   The terms involving $\lambda$ in these equations are playing the physically meaning role doing the effect of the Finsler geometric consideration of the problem.
   \\

\section*{Acknowledgments} FR and SR are thankful to the
Inter-University Centre for Astronomy and Astrophysics (IUCAA),
India for providing Visiting Associateship under which a part of
this work was carried out. FR is also grateful to DST, Govt. of India for financial support under PURSE programme. We are also grateful to the referee for his valuable suggestions.


\begin{thebibliography}{99}

\bibitem{Pais1982} A. Pais, Subtle is the Lord: The Science and
Life of Albert Einstein (Oxford Univ. Press, 1982).

\bibitem{BZ1934} W. Baade and F. Zwicky, Physical Review, {\bf
46}, 76 (1934).

\bibitem{Longair1994} M.S. Longair, High Energy Astrophysics (Vol.
2, Cambridge Univ., p.99, 1994).

\bibitem{Ghosh2007} P. Ghosh, Rotation and Accretion Powered Pulsars (p.2,
World Scientific, 2007).

\bibitem{Ruderman1972} R. Ruderman, Rev. Astr. Astrophys. {\bf 10}, 427
(1972).

\bibitem{gokhroo1994} M.K. Gokhroo, A.L. Mehra, Gen. Relativ. Grav. {\bf 26}, 75
(1994).

\bibitem{Kippenhahn1990} R. Kippenhahn, A. Weigert, Steller Structure and Evolution (Springer-Verlag, 1990).

\bibitem{sokolov1980} A.I. Sokolov, JETP {\bf 79}, 1137 (1980).

\bibitem{sawyer1972} R.F. Sawyer, Phys. Rev. Lett. {\bf 29}, 382 (1972); Erratum Phys. Rev. Lett. {\bf 29}, 823
(1972).

\bibitem{Silva2014} H.O. Silva et al., arXiv: 1411.6286 (2014).

\bibitem{Bowers1974} R.L. Bowers, E.P.T. Liang, Astrophys. J. {\bf 188}, 657
(1917).

\bibitem{Herrera2004} L. Herrera, A. Di Prisco, J. Martin, J. Ospino, N.O. Santos, O. Troconis, Phys. Rev. D {\bf 69}, 084026,
(2004).

\bibitem{Varela2010} V. Varela, F. Rahaman, S. Ray, K. Chakraborty, M. Kalam, Phys. Rev. D {\bf 82}, 044052
(2010).

\bibitem{Rahaman2010a} F. Rahaman, S. Ray, A.K. Jafry, K. Chakraborty, Phys. Rev. D {\bf 82}, 104055
(2010).

\bibitem{Rahaman} F. Rahaman, R. Sharma, S. Ray, R. Maulick, I. Karar, Eur. Phys. J. C {\bf 72}, 2071
(2012).

\bibitem{Rahaman2012a} F. Rahaman, R. Maulick, A.K. Yadav, S. Ray, R. Sharma, Gen. Rel. Grav. {\bf 44}, 107
(2012).

\bibitem{Kalam2012} M. Kalam, F. Rahaman, S. Ray, M. Hossein, I. Karar, J. Naskar, Euro. Phys. J. C {\bf 72}, 2248
(2012).

\bibitem{Hossein2012} Sk. M. Hossein, F. Rahaman, J. Naskar, M. Kalam, S. Ray, Int. J. Mod. Phys. D {\bf 21}, 1250088
(2012).

\bibitem{Kalam2013} M. Kalam, A.A. Usmani, F. Rahaman, S.M. Hossein, I. Karar, R. Sharma, Int. J. Theor. Phys. {\bf 52}, 3319
(2013).

\bibitem{Herrera1997} L. Herrera, N.O. Santos, Phys. Report. {\bf 286}, 53
(1997).

\bibitem{Bao2000} D. Bao, S. S. Chern, and Z. Shen, An Introduction to Riemann–Finsler Geometry, Graduate
Texts in Mathematics (Springer, New York, 2000).

\bibitem{Car}  E. Cartan, Les Espaces de Finsler (Paris, Herman, 1935).

\bibitem{hor}   J. I. Horváth,   Phys. Rev. 80 (1950) 2001.

\bibitem{vac1}   S. Vacaru,   Int. J. Mod. Phys.
D 21 (2012) 1250072; arXiv: 1004.3007.

\bibitem{vac2}   S. Vacaru, Critical remarks on Finsler modifications of gravity and cosmology by Zhe Chang and Xin Li,
Phys. Lett. B 690 (2010) 224-228; arXiv: 1003.0044v2.


  \bibitem{vac3} S. Vacaru,   Nucl. Phys. B, 434 (1997) 590
-656; arXiv: hep-th/9611034.

\bibitem{vac4}  S. Vacaru,   J. Math. Phys. 37 (1996) 508-523.


\bibitem{vac5}  S. Vacaru,   Gener. Relat.
Grav. 44 (2012) 1015-1042; arXiv: 1010.5457.

\bibitem{vac6}   S. Rajpoot and S. Vacaru,  Int. J. Geom. Meth. Mod. Phys. 12 (2015); arXiv: 1506.08696.

\bibitem{vac7}   P. Stavrinos and S. Vacaru,   Class. Quant. Grav. 30 (2013) 055012; arXiv: 1206.3998




\bibitem{Lammerzahl2012} C. L{\"a}mmerzahl, V. Perlick and W. Hasse, Phys. Rev. D, {\bf 86}, 104042 (2012).

\bibitem{Pavlov2010} D.G. Pavlov, AIP Conf. Proc. {\bf 1283}, 180 (2010).

\bibitem{Vacaru2010} S.I. Vacaru, Class. Quantum Gravit. {\bf 27}, 105003 (2010).

\bibitem{Li2014} X. Li and Z. Chang, Phys. Rev. D, {\bf 90}, 064049 (2014).

\bibitem{Akbar1988} H. Akbar-Zadeh, Acad. Roy. Belg. Bull. Cl. Sci., {\bf 74}, 281 (1988).

\bibitem{Rahaman2010} F. Rahaman et al.,  Phys. Rev. D, {\bf 82}, 104055 (2010).

\bibitem{Finch1989} M.R. Finch and J.E.F. Skea, Class. Quantum
Grav., {\bf 6}, 467 (1989).

\bibitem{Lobo2006} F.S.N. Lobo, Class. Quantum Grav., {\bf 23},
1525 (2006).

\bibitem{Mak2003} M.K. Mak and T. Harko, Proc. Roy. Soc. Lond., {\bf
A459}, 393 (2003).

\bibitem{Sharma2007} R. Sharma and S. Maharaj, MNRAS, {\bf 375}, 1265
(2007).

\bibitem{Buchdahl1959} H.A. Buchdahl, Phys. Rev., {\bf 116}, 1027 (1959).

\bibitem{Lindblom1984} L. Lindblom, Astrophys. J., {\bf 278}, 364
(1984).

\bibitem{Herrera1992} L. Herrera, Phys. Lett. A, {\bf 165}, 206 (1992).

\bibitem{Glendenning1997} N.K. Glendenning, Compact Stars: Nuclear Physics,
Particle Physics and General Relativity (Springer-Verlag, New York, p. 70, 1997).

\bibitem{Herjog2011} M. Herjog and F.K. Roepke, arXiv: 1109.0539 [astro-ph.HE] (2014).

\bibitem{Pfeifer2012} C. Pfeifer and M. Wohlfarth, Proceedings of the MG13 Meeting on General Relativity,
Stockholm University, Sweden, 1 - 7 July (2012), doi:
10.1142/9789814623995\_0094


\bibitem{XL}  Xin Li et al,  arxiv: 1309.1758.


 \bibitem{XC} Xin Li $\&$  Z Chang, arXiv:1401.6363.

 \bibitem{Pf}   C.Pfeifer
and M.N.R. Wohlfarth, Phys. Rev. D 85,064009(2012).

\end{thebibliography}
\end{document}